\documentclass[10pt]{article}
\usepackage[letterpaper]{geometry}
\usepackage{hicss}
\usepackage{times}
\usepackage[none]{hyphenat}
\usepackage{url}
\usepackage{latexsym}
\usepackage{minted}
\usepackage{indentfirst}
\usepackage{graphicx}
\graphicspath{{images/}}
\usepackage[
    style=apa,
  ]{biblatex}
\usepackage{amsmath,amssymb,amsthm,epsfig,color,empheq,graphicx,graphics,balance}
\usepackage{enumerate,url,wasysym,epstopdf,enumitem,array}
\usepackage{xcolor}
\usepackage{amsfonts}
\usepackage{multirow}
\usepackage{comment}
\usepackage{booktabs}
\usepackage{accents}

\usepackage{algorithm2e} 
\usepackage{float} 
\usepackage{bbold}
\usepackage{amsmath,amssymb}
\usepackage{booktabs}
\usepackage{tikz}
\usetikzlibrary{arrows.meta,calc,patterns,decorations.pathreplacing}
\usepackage{xcolor}
\title{Probabilistic Repair Logistics Modeling for Utility-Scale PV Inverter Fleets Using Event-Driven Simulation}

\author{Jinlei Wei \\
 Purdue University \\
 {\underline{wei474@purdue.edu}} \\ \\
 \And
  Yongxin Zhang \\
 Texas A\&M University\\
 {\underline{ yongxin25@tamu.edu} } \\ \\
\And
 Guanyu Tian \\
 Texas A\&M University \\
 {\underline{ tiang@tamug.edu} } \\ \\
}
\date{}

\begin{document}
\maketitle
\begin{abstract}
As renewable energy systems expand, inverter availability becomes increasingly important for grid reliability and economics, yet photovoltaic inverter repair logistics remain under-modeled. This paper presents an event-driven Monte Carlo framework for a centralized repair facility with parallel production lines, capturing the full repair cycle from administrative pre-wait and transport to health-driven repair and return-to-inventory. The model incorporates opportunistic scheduling that uses mandatory hold periods to insert additional units onto temporarily idle lines, improving throughput without added capacity. Stage durations are represented by a two-component VaR-style mixture distribution for routine and heavy-tailed delays, while a continuous health score determines repair completion. Calibrated by minimizing the one-dimensional Wasserstein distance between simulated and empirical repair-duration distributions, the model is applied to 43 field-observed repairs, reproducing the empirical bimodal structure with a Wasserstein distance of 53.3 days. Results show that 51.2\% of units are accommodated through opportunistic insertion, indicating that hold periods provide a significant recoverable scheduling resource.
\end{abstract}

\subsubsection*{Keywords: Inverter, repair logistics, stochastic modeling, Wasserstein distance}

\section{Introduction}

Photovoltaic inverters are the most critical and failure-prone power electronic components in utility-scale and distributed renewable energy systems, directly governing the conversion efficiency and grid-connected availability of solar generation assets \parencite{ryan2016assessing, yang2011industry}. As global installed PV capacity has surpassed 1\,TW, even modest improvements in inverter availability translate into substantial revenue and grid reliability benefits: industry data indicate that inverter failures account for the majority of unplanned PV system downtime, with repair durations ranging from days to over a year depending on fault severity, spare-part availability, and logistics constraints \parencite{golnas2013, abdulla2024photovoltaic}. In grid-connected and microgrid contexts, prolonged inverter outages reduce dispatchable renewable capacity, increase reliance on fossil backup generation, and impose direct financial losses through lost feed-in revenue and contractual penalties \parencite{guo2022energy, park2021stochastic}. Effective management of inverter repair logistics is therefore not merely an operational concern but a key determinant of the economic viability and reliability of modern energy systems.

Existing studies have extensively investigated inverter failure modes using field data, accelerated testing, and reliability block diagrams \parencite{roy2024photovoltaic}. Commonly reported dominant failure components include capacitors, power semiconductor modules \parencite{yang2011industry}, heat sinks \parencite{shahzad2019review}, and control boards \parencite{batzelis2016off}. To evaluate and mitigate risk \parencite{karim2025review} in PV systems, researchers and engineers have contributed to operation and maintenance (O\&M) strategies in which statistics, artificial intelligence, and condition monitoring have been introduced to improve fault detection accuracy and efficiency  \parencite{abubakar2021review}.

However, most prior work has focused on failure occurrence, fault diagnosis, and component-level reliability \parencite{nagarajan2019photovoltaic}, rather than on the operational process that occurs after a failed inverter enters a repair facility. The actual situation still lacks comprehensive guidance on coordinating the timing and sequencing of maintenance interventions and on incorporating staff coordination, spare parts, logistics, and supply-chain management into the broader operational context \parencite{abdulla2024photovoltaic}. In conventional reliability, availability, and maintainability analysis, repair is often represented through aggregate indicators such as repair rates \parencite{danjuma2022reliability}, which obscure the multi-stage, resource-constrained, and highly uncertain nature of real repair operations. This simplification is insufficient for inverter repair logistics, where total downtime may include administrative pre-wait, forklift transport, inspection, testing, parts sourcing, repeated repair cycles, re-testing, return handling, and queuing among parallel production lines, each introducing heterogeneous and heavy-tailed delays \parencite{abdulla2024photovoltaic, guo2022energy}. Existing models also fail to capture the interaction between individual repair trajectories and facility-level scheduling rules, although repair-shop studies have shown that scheduling policies affect downtime under limited repair capacity \parencite{liang2013scheduling}, and repair-facility studies have modeled joint inventory and repair scheduling decisions \parencite{ozkan2023joint}.

To bridge these gaps, this paper develops a stochastic, event-driven simulation framework for modeling the end-to-end repair logistics of utility-scale PV inverter fleets at a centralized multi-line repair facility. A health-score-based repair loop is introduced to represent repeated repair interventions and probabilistic outcome transitions. At the facility level, the model incorporates parallel production lines, pending queues, and a three-priority dispatch rule that
allows opportunistic insertion of waiting units during temporary line-release windows. The resulting repair-duration distribution is calibrated and validated against field-observed repair data using the Wasserstein distance. The principal contributions are:
\begin{enumerate}
  \item \textbf{Stochastic simulation model for inverter repair
    logistics.} We develop an event-driven Monte Carlo simulation that reproduces the full repair workflow: multi-stage processing, stochastic health dynamics, and probabilistic branching into \textsc{scrap}, \textsc{decompose}, or \textsc{return}, at the individual unit level.
  \item \textbf{Opportunistic insertion scheduling.}
    We formulate and implement a three-priority dispatch rule that exploits mandatory hold phases to insert additional units onto temporarily idle lines, improving facility throughput without increasing physical capacity.
  \item \textbf{Distribution-level calibration.}
    We minimize the Wasserstein distance between the simulated duration distribution and field observations, providing a
    calibration criterion sensitive to both location and shape, going beyond conventional moment-matching.
\end{enumerate}

%%%%%%%%%%%%%%%%%%%%%%%%%%%%%%%%%%%%%%%%%%%%%%%%%%%%%%%%
\section{Problem Formulation}
\label{sec:formulation}
%%%%%%%%%%%%%%%%%%%%%%%%%%%%%%%%%%%%%%%%%%%%%%%%%%%%%%%%

Photovoltaic (PV) inverters are among the most failure-prone components in large-scale solar plants, and their repair logistics directly determine plant availability and revenue loss during downtime. Unlike consumable parts, a faulty inverter typically undergoes a structured multi-stage assessment before a return-to-service or decommissioning decision is reached. This process involves physical inspection, functional testing, one or more active repair interventions, and final re-testing, each of which carries substantial and heterogeneous time uncertainty.

This paper models the end-to-end repair workflow of a fleet of $M$ inverter units processed at a centralized repair facility equipped with $N$ parallel production lines. The primary quantity of interest is the \emph{repair duration} $D_i$ of each unit $i$---defined as the total elapsed time from the moment a unit enters the business pipeline to the moment it either returns to the plant inventory or is decommissioned. Because $D_i$ is driven by a cascade of stochastic events, we adopt a Monte Carlo simulation framework that explicitly models every sub-process and scheduling interaction, then validates the resulting duration distribution against field-observed repair durations using the Wasserstein distance.

Two operational features distinguish the facility studied here from standard single-server queues. First, units arrive sequentially with random inter-arrival gaps and may wait in a pending queue if all lines are occupied; the time a unit actually enters a line depends on both the arrival schedule and real-time line availability. Second, during a mandatory intermediate hold phase (\emph{wait\,2}), the production line is temporarily relinquished and may be used by another waiting unit---at most one per hold window. This \emph{opportunistic insertion} rule improves throughput without adding physical capacity.

\begin{figure}[ht]
  \centering
  \includegraphics[width=0.8\linewidth]{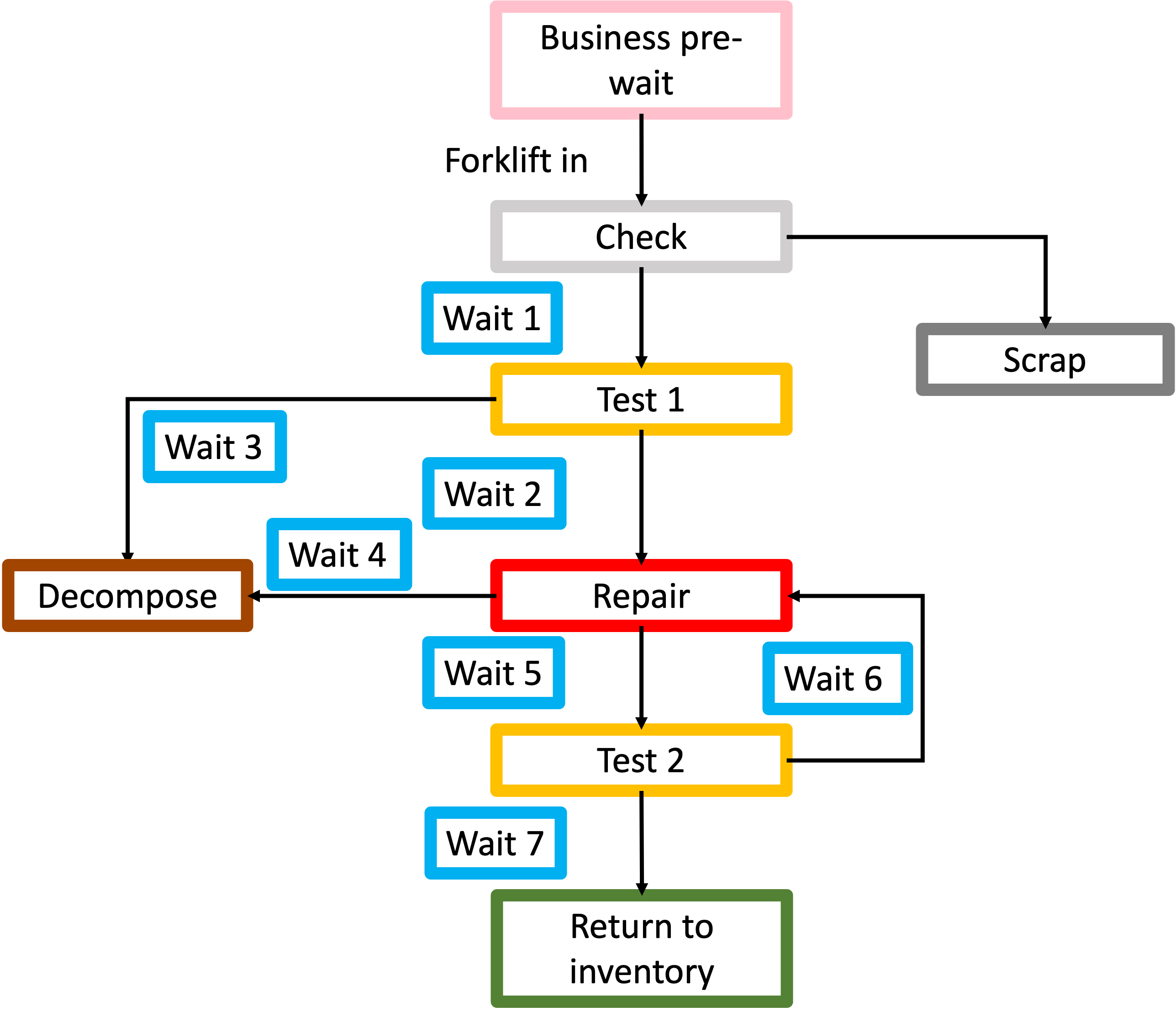}
  \caption{Single inverter repair workflow.}
  \label{fig:workflow}
\end{figure}

Figure~\ref{fig:workflow} illustrates the branching structure of the repair workflow. Upon entering a production line, every unit passes through \textbf{Check}; a fraction $p_s$ is immediately condemned (\textsc{scrap}), while the remainder proceeds to \textbf{Wait\,1} and \textbf{Test\,1}. Units assessed as not requiring active repair (probability
$1 - p_r$) are routed via \textbf{Wait\,3} to \textbf{Decompose}, whereas units entering the repair loop cycle through \textbf{Wait\,2} $\to$ \textbf{Repair} $\to$ \textbf{Wait\,5} $\to$ \textbf{Test\,2} until the health score either reaches the recovery threshold (\textbf{Wait\,7} $\to$ \textsc{return to inventory}) or falls below the minimum threshold (\textbf{Wait\,4} $\to$ \textbf{Decompose}). Three terminal outcomes are therefore possible: \textsc{scrap}, \textsc{decompose}, and \textsc{return}.

Figure~\ref{fig:timeline_example} then illustrates one specific realization of this workflow, the \textsc{return} outcome, from the perspective of wall-clock time, showing how each stage contributes to the total repair duration $D_i$.

% ------------------------------------------------------------------
\definecolor{cpink}  {RGB}{255,182,193}
\definecolor{cpurple}{RGB}{160,100,200}
\definecolor{cgray}  {RGB}{180,180,180}
\definecolor{cblue}  {RGB}{173,216,230}
\definecolor{corange}{RGB}{255,165, 80}
\definecolor{cred}   {RGB}{220, 80, 80}
\definecolor{cbrown} {RGB}{160, 82, 45}
\definecolor{cgreen} {RGB}{ 80,160, 80}

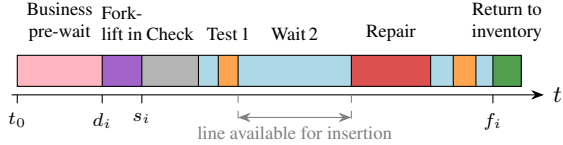
\begin{figure}[ht]
\centering
\begin{tikzpicture}[
    >=Stealth, font=\small,
    lbl/.style={font=\scriptsize, align=center},
    scale=0.75
]
\def\h{0.55}\def\y{0}
\pgfmathsetmacro{\ylo}{\y-\h/2}
\pgfmathsetmacro{\yhi}{\y+\h/2}

\fill[cpink]   (0.0,\ylo) rectangle (1.5,\yhi);
\fill[cpurple] (1.5,\ylo) rectangle (2.2,\yhi);
\fill[cgray]   (2.2,\ylo) rectangle (3.2,\yhi);
\fill[cblue]   (3.2,\ylo) rectangle (3.55,\yhi);
\fill[corange] (3.55,\ylo) rectangle (3.9,\yhi);
\fill[cblue]   (3.9,\ylo) rectangle (5.9,\yhi);
\fill[cred]    (5.9,\ylo) rectangle (7.3,\yhi);
\fill[cblue]   (7.3,\ylo) rectangle (7.7,\yhi);
\fill[corange] (7.7,\ylo) rectangle (8.1,\yhi);
\fill[cblue]   (8.1,\ylo) rectangle (8.4,\yhi);
\fill[cgreen]  (8.4,\ylo) rectangle (8.9,\yhi);

\draw (0.0,\ylo) rectangle (8.9,\yhi);
\foreach \x in {1.5,2.2,3.2,3.55,3.9,5.9,7.3,7.7,8.1,8.4}{
  \draw (\x,\ylo) -- (\x,\yhi);
}
\draw[->] (-0.1,\ylo-0.15) -- (9.3,\ylo-0.15) node[right]{$t$};
\foreach \x/\lab in {0.0/$t_0$, 1.5/$d_i$, 2.2/$s_i$, 8.4/$f_i$}{
  \draw (\x,\ylo-0.15) -- (\x,\ylo-0.30);
  \node[below, font=\scriptsize] at (\x,\ylo-0.30) {\lab};
}
\node[lbl, above=2pt] at (0.75,\yhi) {Business\\pre-wait};
\node[lbl, above=2pt] at (1.85,\yhi) {Fork-\\lift in};
\node[lbl, above=2pt] at (2.70,\yhi) {Check};
\node[lbl, above=2pt] at (3.73,\yhi) {Test\,1};
\node[lbl, above=2pt] at (4.90,\yhi) {Wait\,2};
\node[lbl, above=2pt] at (6.60,\yhi) {Repair};
\node[lbl, above=2pt] at (8.65,\yhi) {Return to\\inventory};
\draw[dashed, gray] (3.9,\ylo-0.15) -- (3.9,\ylo-0.60);
\draw[dashed, gray] (5.9,\ylo-0.15) -- (5.9,\ylo-0.60);
\draw[<->, gray] (3.9,\ylo-0.55) -- (5.9,\ylo-0.55)
  node[midway, below, font=\scriptsize, gray]{line available for insertion};
\end{tikzpicture}
\caption{Repair timeline illustration for a single inverter unit with an outcome of \textsc{return}.
%   The bar spans $D_i$ from business pre-wait start
%   ($t_0 = d_i - w_i$) to return-to-inventory completion.
%   During \emph{wait\,2} the production line is released and may
%   accept one inserted unit from the pending queue.
%   Colours:
%   \textcolor{cpink}{\rule{6pt}{6pt}}~business pre-wait,
%   \textcolor{cpurple}{\rule{6pt}{6pt}}~forklift transport,
%   \textcolor{cgray}{\rule{6pt}{6pt}}~check,
%   \textcolor{cblue}{\rule{6pt}{6pt}}~waiting/hold stages,
%   \textcolor{corange}{\rule{6pt}{6pt}}~test stages,
%   \textcolor{cred}{\rule{6pt}{6pt}}~repair,
%   \textcolor{cgreen}{\rule{6pt}{6pt}}~return to inventory.%
}
\label{fig:timeline_example}
\end{figure}
% ------------------------------------------------------------------

The following phases describe each segment of $D_i$:

\begin{itemize}
  \item \textbf{Business pre-wait} $[t_0,\, d_i]$: An administrative holding period $w_i$ covering documentation, procurement approval, and logistics clearance, which does not consume any production line capacity. The unit is dispatched at time $d_i = t_0 + w_i$.
  \item \textbf{Inbound forklift transport} $[d_i,\, s_i]$: The unit is transported from the staging area to the assigned production line with duration $\tau_i^{\mathrm{in}}$, arriving at $s_i = d_i + \tau_i^{\mathrm{in}}$.
  \item \textbf{Check}: An initial physical inspection of duration $\tau^{\mathrm{chk}}$. With probability $p_s$ the unit is immediately condemned (\textsc{scrap}); otherwise it proceeds to testing.
  \item \textbf{Wait\,1 $\to$ Test\,1}: A preparatory hold followed by a functional test. With probability $p_r$ the unit requires active repair and enters the repair loop, where $p_r$ denotes the probability of a repair-needed assessment; otherwise, it proceeds via Wait\,3 to decomposition (\textsc{decompose}).
  \item \textbf{Wait\,2}: A mandatory hold during which components are sourced or soaked. The production line is \emph{released} for the entire duration of this phase, creating an opportunity for one additional unit to be processed.
  \item \textbf{Repair}: Active maintenance of duration $\tau^{\mathrm{rep}}$. After each cycle the unit's health score     $h$ is incremented; if $h$ reaches the recovery threshold $h^*$ the unit is cleared for return, if $h$ falls below $h_{\min}$ it is condemned to decomposition, otherwise a further repair cycle follows.
  \item \textbf{Return to inventory} $[f_i,\, f_i + \tau_i^{\mathrm{out}}]$ (\textsc{return} only): The repaired unit is transported back to the plant inventory with duration $\tau_i^{\mathrm{out}}$.
\end{itemize}

The total repair duration encompasses every phase:
\begin{equation}
  D_i \;=\; \bigl(f_i + \tau_i^{\mathrm{out}}\bigr) - t_0
        \;=\; \bigl(f_i + \tau_i^{\mathrm{out}}\bigr)
              - \bigl(d_i - w_i\bigr).
  \label{eq:repair_duration}
\end{equation}

%%%%%%%%%%%%%%%%%%%%%%%%%%%%%%%%%%%%%%%%%%%%%%%%%%%%%%%%
\section{Probabilistic Modeling for Inverter Repair Logistics}
\label{sec:approach}
%%%%%%%%%%%%%%%%%%%%%%%%%%%%%%%%%%%%%%%%%%%%%%%%%%%%%%%%

This section formalizes the key modeling components that underpin the simulation: the stage duration distributions, the health-score dynamics governing repair loop termination, and the dispatch priority rules that govern line assignment.

\subsection{Stage Duration Distributions}
\label{sec:distributions}

Field records of inverter repair operations exhibit a characteristic two-regime pattern: the majority of interventions at each stage are completed within a predictable range, yet a non-negligible fraction incur substantially longer delays due to part unavailability, logistics disruption, or repeated diagnostic cycles. A single parametric family cannot simultaneously capture both the concentrated bulk of routine cases and the heavy right tail of exceptional ones without severely distorting one or the other.

To reflect this structure, each processing stage $s$ is characterized by three quantile parameters $(v_{50}^{(s)},\, v_{80}^{(s)},\, v_{100}^{(s)})$---the median, 80th-percentile, and practical maximum---elicited directly from historical records without committing to a specific distributional family. Stage durations are drawn from the two-component mixture:
\begin{equation}
  \tau^{(s)} \sim
  \begin{cases}
    \mathrm{clip}\!\left(
      \mathcal{N}\!\Bigl(v_{50}^{(s)},\;
      \sigma_s^2\Bigr),\;
      0,\; v_{100}^{(s)}\right)
      & \text{w.p. } \pi, \\[6pt]
    \exp\!\Bigl(
      \mathcal{U}\bigl(\ln v_{80}^{(s)},\;\ln v_{100}^{(s)}\bigr)
    \Bigr)
      & \text{w.p. } 1 - \pi,
  \end{cases}
  \label{eq:var_sampler}
\end{equation}
where $\sigma_s = \max(0.15\,v_{50}^{(s)},\, 0.2)$ and $\pi \in (0,1)$ is the mixture weight. The normal component captures routine operations; the log-uniform component produces a slowly decaying heavy tail consistent with field observations. Stage durations are treated as continuous random variables measured in days, consistent with standard practice in stochastic maintenance modeling \parencite{jardine2013, marseguerra2000}. Continuous-valued repair times naturally arise when durations represent elapsed calendar time accumulated across logistics, waiting, and active work sub-phases, each of which contributes a non-integer increment; integer-day reporting in field records reflects administrative rounding rather than an inherent discreteness of the underlying process.

\subsection{Health-Score Dynamics}
\label{sec:health}

Rather than modeling repair success as a single Bernoulli trial, we track a continuous health score $h \in [0,1]$ that accumulates the effect of successive repair interventions. The initial health score is modeled as a clipped normal distribution to reflect the empirical observation that most arriving units retain partial functionality, with the upper clip at 0.85, ensuring that no unit enters the repair loop already near full recovery. The initial score is drawn as:
\begin{equation}
  h_0 = \mathrm{clip}\!\left(
        \mathcal{N}(\mu_h, \sigma_h^2),\; 0,\; 0.85\right).
  \label{eq:health_init}
\end{equation}
After each repair cycle the score is updated by a fixed increment:
\begin{equation}
  h_{k+1} = \min\!\left(1,\; h_k + \Delta h\right).
  \label{eq:health_update}
\end{equation}
The repair loop terminates when:
\begin{equation}
  \text{outcome} =
  \begin{cases}
    \textsc{return}    & \text{if } h_k \geq h^*, \\
    \textsc{decompose} & \text{if } h_k < h_{\min}.
  \end{cases}
  \label{eq:health_termination}
\end{equation}
If neither condition is satisfied the loop continues to the next iteration. Because $\Delta h$ is fixed, the loop terminates in at most $\lceil(1 - h_0)/\Delta h\rceil$ iterations in the \textsc{return} direction, bounding the processing time. This formulation is directly interpretable in terms of observable inverter condition metrics and naturally represents multi-cycle repairs with diminishing returns.

\subsection{Dispatch Priority Rules}
\label{sec:dispatch}

At each dispatch decision, a production line is assigned to the incoming unit according to the following three-level priority hierarchy:
\begin{itemize}
  \item[\textbf{P1}] \textbf{Immediate dispatch}: if any line is currently idle, the unit is dispatched at once to the earliest available line.
  \item[\textbf{P2}] \textbf{Wait\,2 insertion}: if no line is idle but at least one line is in an open wait\,2 window with sufficient remaining time for the unit to arrive before the window closes, the unit is inserted into that window; at most one unit may be inserted per window, preserving the constraint that a single line hosts only one active process at a time.
  \item[\textbf{P3}] \textbf{Deferred dispatch}: if neither P1 nor P2 applies, the unit waits in a pending queue until the
    earliest line becomes free.
\end{itemize}
All stochastic quantities for every unit are pre-sampled once before the scheduling loop begins, using a seeded generator (\texttt{numpy.random.default\_rng}). This ensures that simulation outcomes are independent of the scheduling policy, enabling fair comparison of alternative dispatch strategies on identical realizations.

%%%%%%%%%%%%%%%%%%%%%%%%%%%%%%%%%%%%%%%%%%%%%%%%%%%%%%%%
\section{Model Calibration and Evaluation}
\label{sec:metric}
%%%%%%%%%%%%%%%%%%%%%%%%%%%%%%%%%%%%%%%%%%%%%%%%%%%%%%%%

\subsection{Wasserstein Distance}

Assessing the fidelity of a stochastic simulation requires a goodness-of-fit measure that is sensitive to the full shape of the output distribution, not merely its first two moments. This requirement is particularly stringent in the present setting, where the empirical repair duration distribution is bimodal and spans two orders of magnitude: a mean-variance criterion would conflate units with very short durations (scrapped after check) with those requiring multi-cycle repair, masking structural mismatches between simulated and observed distributions. 

We therefore adopt the one-dimensional Wasserstein distance $W_1$, also known as the Earth Mover's Distance, as the primary goodness-of-fit criterion. Unlike the Kolmogorov-Smirnov statistic, which is sensitive only to the point of maximum CDF discrepancy, $W_1$ integrates the distributional difference over the entire support, penalizing both location shifts and shape mismatches simultaneously. Let $\hat{F}_M$ be the empirical CDF of the $M$ simulated repair durations $\{D_i\}$ and $F^*_M$ the empirical CDF of the $M$ field-observed durations $\{d^*_j\}$. Then:
\begin{equation}
\begin{aligned}
  W_1(\hat{F}_M,\, F^*_M)
  &= \int_{-\infty}^{\infty}
     \bigl|\hat{F}_M(x) - F^*_M(x)\bigr|\,\mathrm{d}x \\
  &= \int_0^1
     \bigl|\hat{F}_M^{-1}(u) - {F^*_M}^{-1}(u)\bigr|\,\mathrm{d}u.
\end{aligned}
\label{eq:wasserstein}
\end{equation}
The quantile representation makes the operational interpretation transparent: $W_1$ is the average absolute difference in repair duration between the simulated and empirical distributions at each quantile level, measured in days. Since both $\hat{F}_M$ and $F^*_M$ are empirical step functions derived from equal-sized samples of $M = 43$ observations, \eqref{eq:wasserstein} reduces to a closed-form expression in terms of the sorted sample vectors $D_{(1)} \leq \cdots \leq D_{(M)}$ and $d^*_{(1)} \leq \cdots \leq d^*_{(M)}$:
\begin{equation}
  W_1(\hat{F}_M,\, F^*_M)
  = \frac{1}{M} \sum_{k=1}^{M} \bigl|D_{(k)} - d^*_{(k)}\bigr|.
  \label{eq:w1_discrete}
\end{equation}

\subsection{Calibration Objective}

The simulated repair duration $D_i$, as defined in \eqref{eq:repair_duration}, is computed for each unit $i$ as the full elapsed time from the start of the business pre-wait to the end of the return-to-inventory trip (or to the terminal processing event for \textsc{scrap} and \textsc{decompose} outcomes). The collection $\{D_i\}_{i=1}^{M}$ forms the simulated sample against which the empirical dataset $\{d^*_j\}_{j=1}^{M}$ is compared.

Model calibration is framed as the minimization of $W_1$ over the set of VaR parameters $\boldsymbol{\theta}$, which includes the stage-level quantiles $\{(v_{50}^{(s)}, v_{80}^{(s)}, v_{100}^{(s)})\}$, the mixture weight $\pi$, and the health dynamics $(\mu_h, \sigma_h, \Delta h, h_{\min}, h^*)$:
\begin{equation}
  \min_{\boldsymbol{\theta}} \;
  W_1\!\Bigl(
    \hat{F}_M\!\left(\{D_i(\boldsymbol{\theta})\}\right),\;
    F^*_M\!\left(\{d^*_j\}\right)
  \Bigr).
  \label{eq:calibration}
\end{equation}
This formulation goes beyond conventional moment-matching: rather than requiring only that the simulated mean matches the empirical value, \eqref{eq:calibration} requires the entire simulated duration distribution to be close to the empirical distribution in the transport sense, capturing the shape and relative weight of both the short-duration and long-duration clusters. The calibrated parameter values and the resulting $W_1$ are reported in Section~\ref{sec:test}.

%%%%%%%%%%%%%%%%%%%%%%%%%%%%%%%%%%%%%%%%%%%%%%
\section{Numerical Tests}
\label{sec:test}
%%%%%%%%%%%%%%%%%%%%%%%%%%%%%%%%%%%%%%%%%%%%%%

This section presents the numerical study in three parts: an analysis of the empirical repair duration dataset (\S\ref{sec:empirical}), the parameter specification and calibration procedure (\S\ref{sec:params}), and the simulation results with distributional validation (\S\ref{sec:results}).

%%%%%%%%%%%%%%%%%%%%%%%%%%%%%%%%%%%%%%%%%%%%%%
\subsection{Empirical Data Analysis}
\label{sec:empirical}
%%%%%%%%%%%%%%%%%%%%%%%%%%%%%%%%%%%%%%%%%%%%%%

The empirical dataset $\mathbf{d}^*$ comprises $M = 43$ field-observed repair durations for PV inverter units, listed in Table~\ref{tab:empirical_data}. Summary statistics are reported alongside the raw data; the mean is 216.0 days with a standard deviation of 181.2 days, indicating high variability relative to the mean.

\begin{table}[ht]
\centering
\small
\caption{Empirical repair duration dataset $\mathbf{d}^*$ (days), in original observation order, with summary statistics.}
\label{tab:empirical_data}
\begin{tabular}{@{}rrrrrrrrr@{}}
\toprule
608 & 690 & 531 & 407 & 498 & 434 & 401 & 263 & 246 \\
315 & 313 & 360 & 367 & 320 & 315 & 308 & 365 & 134 \\
 51 & 169 & 215 & 176 & 170 & 170 & 170 & 132 &  62 \\
122 &   1 & 135 &  94 &  15 &  10 &   5 &  11 &  46 \\
411 & 116 & 107 &   7 &   9 &   8 &   1 & \multicolumn{2}{c}{} \\
\midrule
\multicolumn{4}{l}{Mean $\;/\;$ Median}
  & \multicolumn{5}{r}{216.0 $\;/\;$ 170.0 days} \\
\multicolumn{4}{l}{Std.\ deviation}
  & \multicolumn{5}{r}{181.2 days} \\
\multicolumn{4}{l}{Min $\;/\;$ Max}
  & \multicolumn{5}{r}{1 $\;/\;$ 690 days} \\
\bottomrule
\end{tabular}
\end{table}

\begin{figure}[ht]
  \centering
  \includegraphics[width=0.85\linewidth]{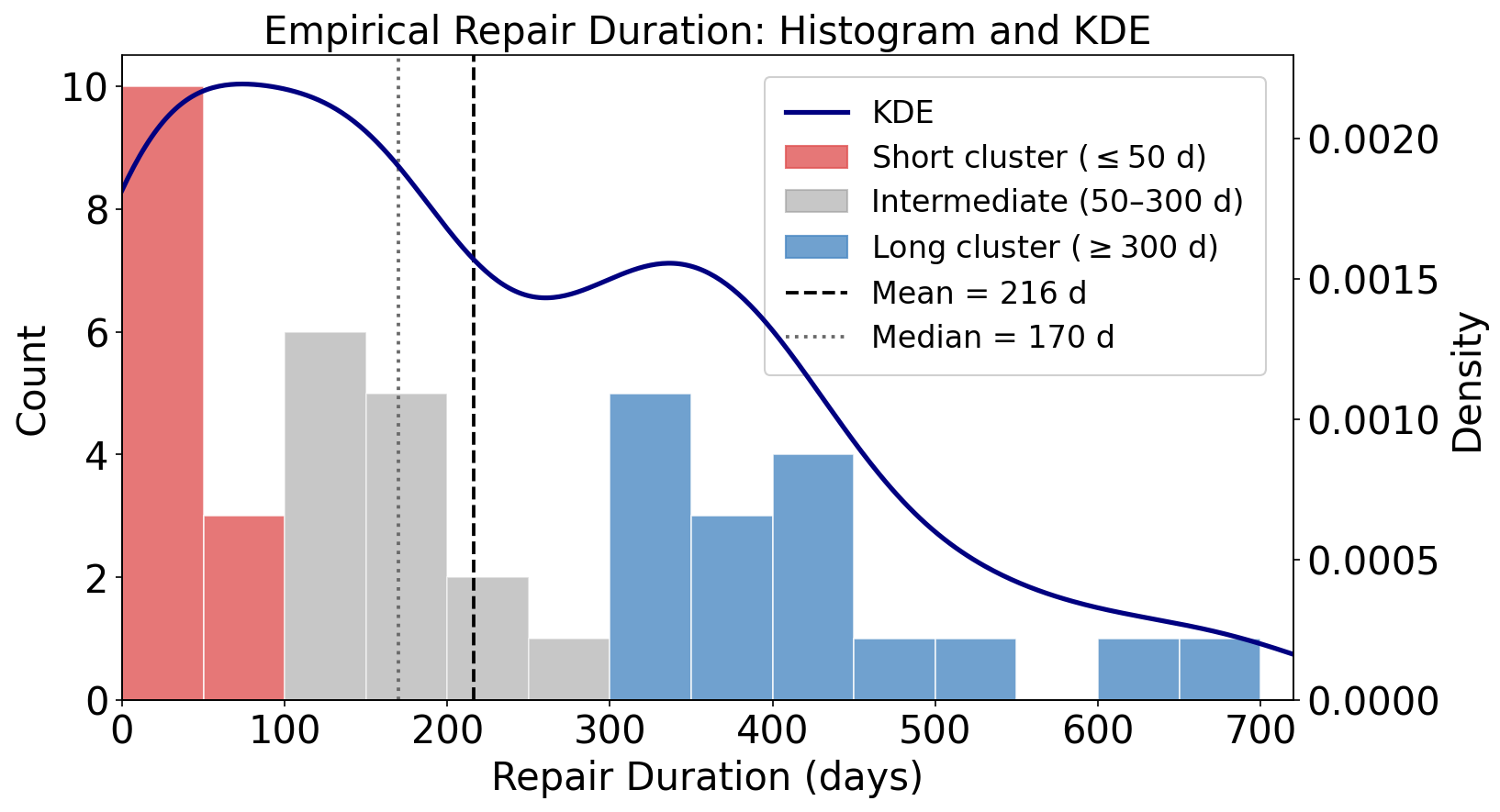}
  \caption{Histogram of the 43 empirical inverters' repair durations.}
  \label{fig:empirical_hist}
\end{figure}

Figure~\ref{fig:empirical_hist} presents the histogram with KDE overlay of the empirical dataset. The distribution is markedly bimodal. A \emph{short-duration cluster} ($\leq 50$ days, red) accounts for units condemned after the initial check (\textsc{scrap}) or quickly repaired. A \emph{long-duration cluster} ($\geq 300$ days, blue) corresponds to units that underwent one or more active repair cycles, with durations accumulating across wait\,2 holds, repair interventions,
and post-repair testing. The intermediate region (50--300 days, grey) represents units resolved via limited repair cycles or early-stage decomposition.

\begin{figure}[ht]
  \centering
  \includegraphics[width=0.85\linewidth]{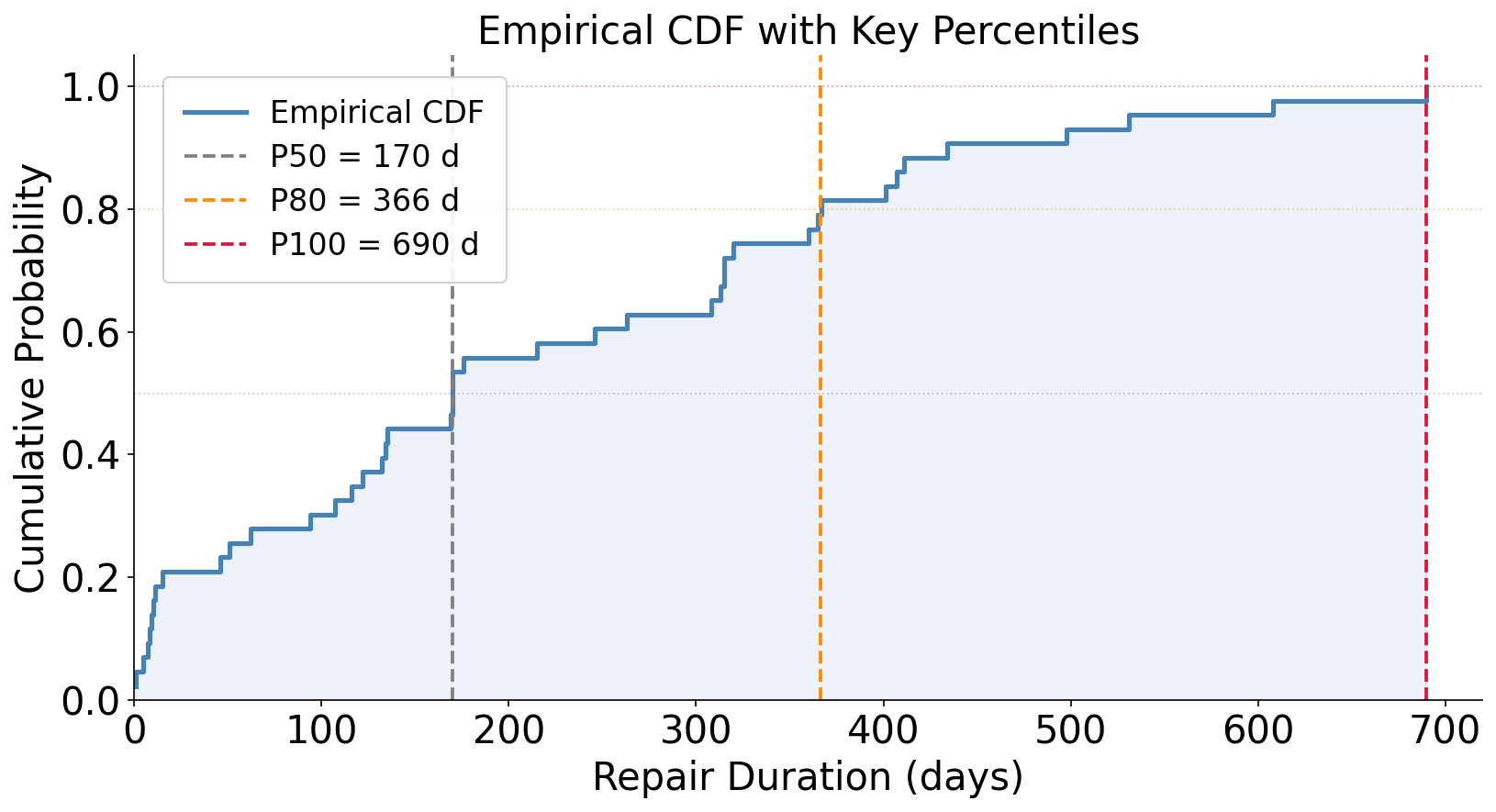}
  \caption{Empirical CDF of the 43 repair durations.}
  \label{fig:empirical_cdf}
\end{figure}

Figure~\ref{fig:empirical_cdf} shows the empirical CDF with key percentile markers. The curve rises steeply below 50 days (approximately 25\% of observations), flattens in the intermediate range, and resumes a steady climb between 100 and 700 days. The 50th, 80th, and 100th percentiles are 170, 401, and 690 days respectively, confirming a heavy upper tail where a small number of units account for a disproportionate share of total repair time. These structural features directly motivate the two-component VaR mixture in \eqref{eq:var_sampler} and the health-score dynamics in \eqref{eq:health_init}--\eqref{eq:health_termination}.

%%%%%%%%%%%%%%%%%%%%%%%%%%%%%%%%%%%%%%%%%%%%%%
\subsection{Parameter Specification and Calibration}
\label{sec:params}
%%%%%%%%%%%%%%%%%%%%%%%%%%%%%%%%%%%%%%%%%%%%%%

\subsubsection{Calibration Procedure}

Parameter values were determined in two stages. In the first stage, initial estimates were derived directly from the empirical dataset. The VaR quantiles $(v_{50}^{(s)}, v_{80}^{(s)}, v_{100}^{(s)})$ for each processing stage were set to reflect the observed percentiles of the overall duration distribution, apportioned across stages according to domain knowledge of typical inverter repair workflows. The health-score parameters $(\mu_h, \sigma_h)$ were chosen to reflect the empirical return rate (approximately 70\% of units in the field are successfully repaired), and the health increment $\Delta h$ was set so that a unit with an average initial health score requires at most two repair cycles to reach the recovery
threshold. 

In the second stage, the parameters were refined by iteratively running the simulation and observing $W_1(\hat{F}_M, F^*_M)$ defined in \eqref{eq:wasserstein}. Adjustments were guided by the direction of the distributional mismatch: when the simulated mean exceeded the empirical mean, the upper VaR parameters $v_{80}$ and $v_{100}$ of the most time-intensive stages (Wait\,2 and Repair) were reduced; when the simulated short-duration cluster was too sparse, the scrap probability $p_s$ or the health threshold $h_{\min}$ was adjusted. This iterative procedure converged to the calibrated parameter set reported below, achieving $W_1 = 53.3$ days.

\subsubsection{Calibrated Parameters}

The stage VaR parameters follow the mixture distribution in \eqref{eq:var_sampler} with mixture weight $\pi = 0.8$; calibrated values are listed in Table~\ref{tab:params}. The health-score parameters follow \eqref{eq:health_init}--\eqref{eq:health_termination} with the calibrated values $\mu_h = 0.70$, $\sigma_h = 0.30$, $\Delta h = 0.80$, $h_{\min} = 0.90$, and $h^* = 0.999$. All scalar parameters are consolidated in Table~\ref{tab:summary}.

\begin{table}[ht]
\centering
\small
\caption{VaR-style duration parameters for each processing stage (days).}
\label{tab:params}
\begin{tabular}{@{}lccc@{}}
\toprule
\textbf{Stage} & $v_{50}$ & $v_{80}$ & $v_{100}$ \\
\midrule
Check     & 10 &  20 &  40 \\
Scrap     &  5 &  10 &  20 \\
Wait\,1   & 10 &  20 &  40 \\
Test\,1   & 10 &  20 &  40 \\
Wait\,2   & 30 & 100 & 200 \\
Repair    & 30 & 100 & 200 \\
Wait\,3   & 10 &  40 &  90 \\
Wait\,4   & 10 &  40 &  90 \\
Wait\,5   & 10 &  20 &  40 \\
Test\,2   & 10 &  20 &  40 \\
Wait\,6   &  3 &  15 &  30 \\
Wait\,7   &  3 &  15 &  30 \\
Decompose & 30 & 100 & 200 \\
\bottomrule
\end{tabular}
\end{table}
\begin{table}[ht]
\centering
\small
\caption{Summary of model parameters.}
\label{tab:summary}
\begin{tabular}{@{}lll@{}}
\toprule
\textbf{Parameter} & \textbf{Value} & \textbf{Description} \\
\midrule
$M$             & 43       & Number of units simulated \\
$N$             & 3        & Number of production lines \\
$\mu_\delta$    & 30 days  & Mean inter-arrival interval \\
$\sigma_\delta$ & 5 days   & Std.\ of inter-arrival interval \\
$\mu_f$         & 2 days   & Mean forklift travel time \\
$\sigma_f$      & 0.5 days & Std.\ of forklift travel time \\
$\mu_w$         & 7 days   & Mean business pre-wait \\
$\sigma_w$      & 3 days   & Std.\ of business pre-wait \\
$p_s$           & 0.08     & Scrap probability \\
$p_r$           & 0.90     & Repair-needed probability \\
$\mu_h$         & 0.70     & Mean initial health score \\
$\sigma_h$      & 0.30     & Std.\ of initial health score \\
$\Delta h$      & 0.80     & Health increment per repair \\
$h_{\min}$      & 0.90     & Decompose threshold \\
$h^*$           & 0.999    & Recovery threshold \\
$\pi$           & 0.8      & Normal component mixture weight \\
$c_{\max}$      & 1        & Max insertions per wait2 window \\
\bottomrule
\end{tabular}
\end{table}

%%%%%%%%%%%%%%%%%%%%%%%%%%%%%%%%%%%%%%%%%%%%%%
\subsection{Results}
\label{sec:results}
%%%%%%%%%%%%%%%%%%%%%%%%%%%%%%%%%%%%%%%%%%%%%%

The simulation is run with 43 units and 3 production lines using the random number generator of Python Numpy package with a fixed seed, with all stochastic quantities pre-sampled as described in \S\ref{sec:dispatch}.

\subsubsection{Distribution Comparison and Wasserstein Distance}

\begin{figure}[ht]
  \centering
  \includegraphics[width=\linewidth]{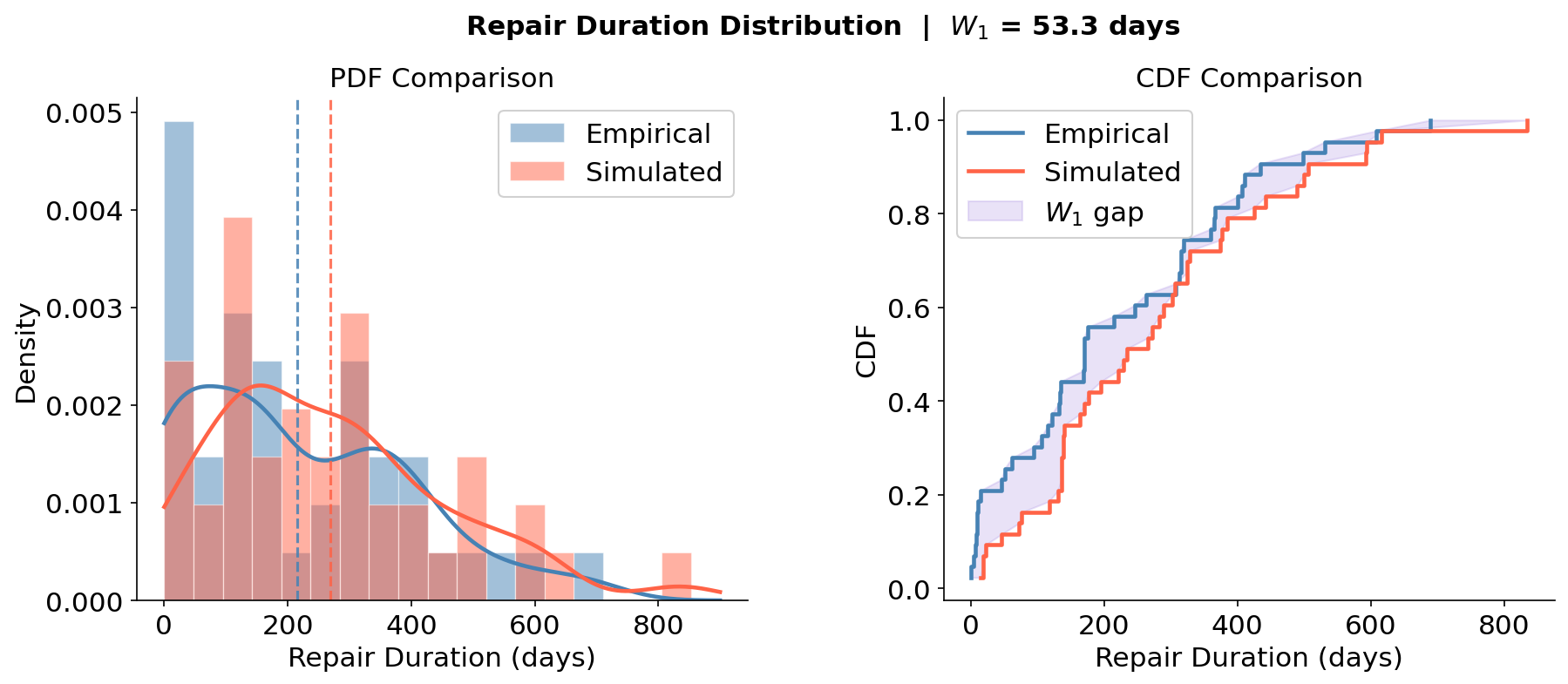}
  \caption{Left: PDF comparison between simulated and empirical
           repair durations, with KDE curves and mean lines.
           Right: CDF comparison; the shaded region represents
           the integrand of $W_1 = 53.3$ days.}
  \label{fig:distribution}
\end{figure}

Fig.~\ref{fig:distribution} compares the simulated and empirical repair duration distributions via both the PDF and the CDF. Both distributions exhibit the bimodal structure identified in \S\ref{sec:empirical}: a short-duration cluster from units resolved through scrap or single-cycle repair, and a long-duration cluster from units requiring multiple repair cycles. The simulated distribution reproduces this two-regime shape with a mean of $\bar{D}_{\rm sim} = 269.3$ days, compared to the empirical mean $\bar{d}^* = 216.0$ days, yielding $W_1 = 53.3$ days. Relative to the empirical standard deviation of 181.2 days and range of 1--690 days, this represents a moderate fit: the overall shape and bimodal character are well captured, while the residual upward bias suggests that the upper VaR parameters of Wait\,2 and Repair could be tightened in further calibration rounds. The shaded region in the CDF panel directly visualizes the $W_1$ gap.

\subsubsection{Per-Unit Repair Timeline}

\begin{figure}[ht]
  \centering
  \includegraphics[width=\linewidth]{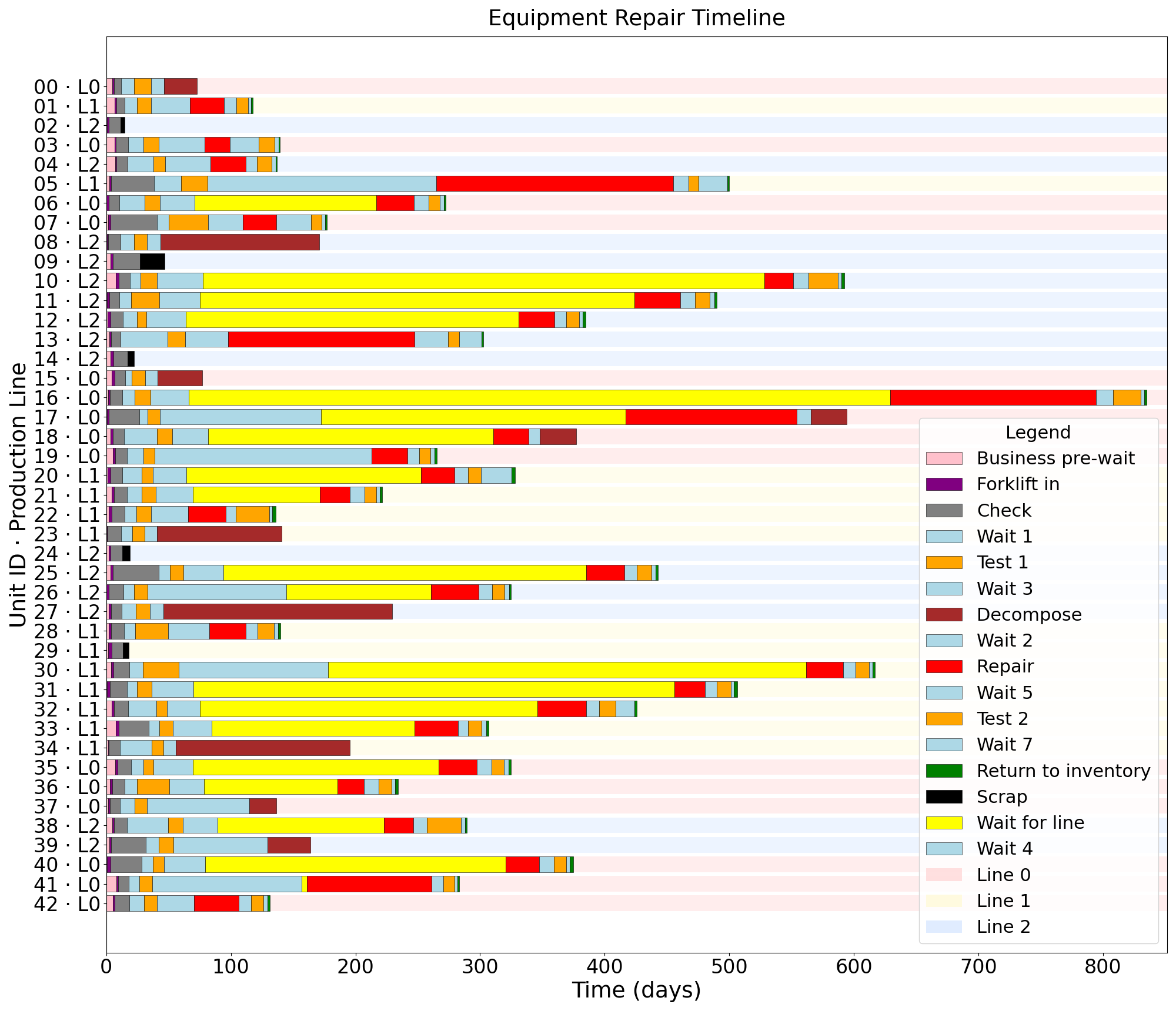}
  \caption{Per-unit timeline comparison for all 43 simulated units.}
  \label{fig:per_unit}
\end{figure}

Fig.~\ref{fig:per_unit} shows the per-unit compact timeline for all 43 simulated units, with each bar starting from the unit's own dispatch time and extending through every processing stage until the terminal outcome. Background shading identifies the production line assignment (red: Line~0; yellow: Line~1; blue: Line~2). The prominent yellow segments indicate periods in which a unit has completed its wait\,2 hold but must wait for the line to be released by a previously inserted unit; this delay arises directly from the P2 insertion rule described in \S\ref{sec:dispatch}, where the inserted unit may still be processing when the host unit's wait\,2 window closes. Several observations are noteworthy: 
\begin{itemize}
  \item Units with a \textsc{scrap} outcome (e.g., Units~02, 09, 14, 24, 29) exhibit negligible total durations, as processing terminates immediately after the check stage.
  \item The longest per-unit durations arise from extended \texttt{wait-for-line} periods, most prominently in Units~10, 16, 30, and 31, where the line remained occupied by an inserted unit well beyond the scheduled wait\,2 window.
  \item Wait\,2 and repair stages dominate the timeline for multi-cycle units, confirming that the tail of the duration distribution is driven by repeated repair iterations rather
    than administrative delays.
\end{itemize}

\subsubsection{Outcome Breakdown and Dispatch Slot Distribution}

\begin{figure}[ht]
  \centering
  \includegraphics[width=\linewidth]{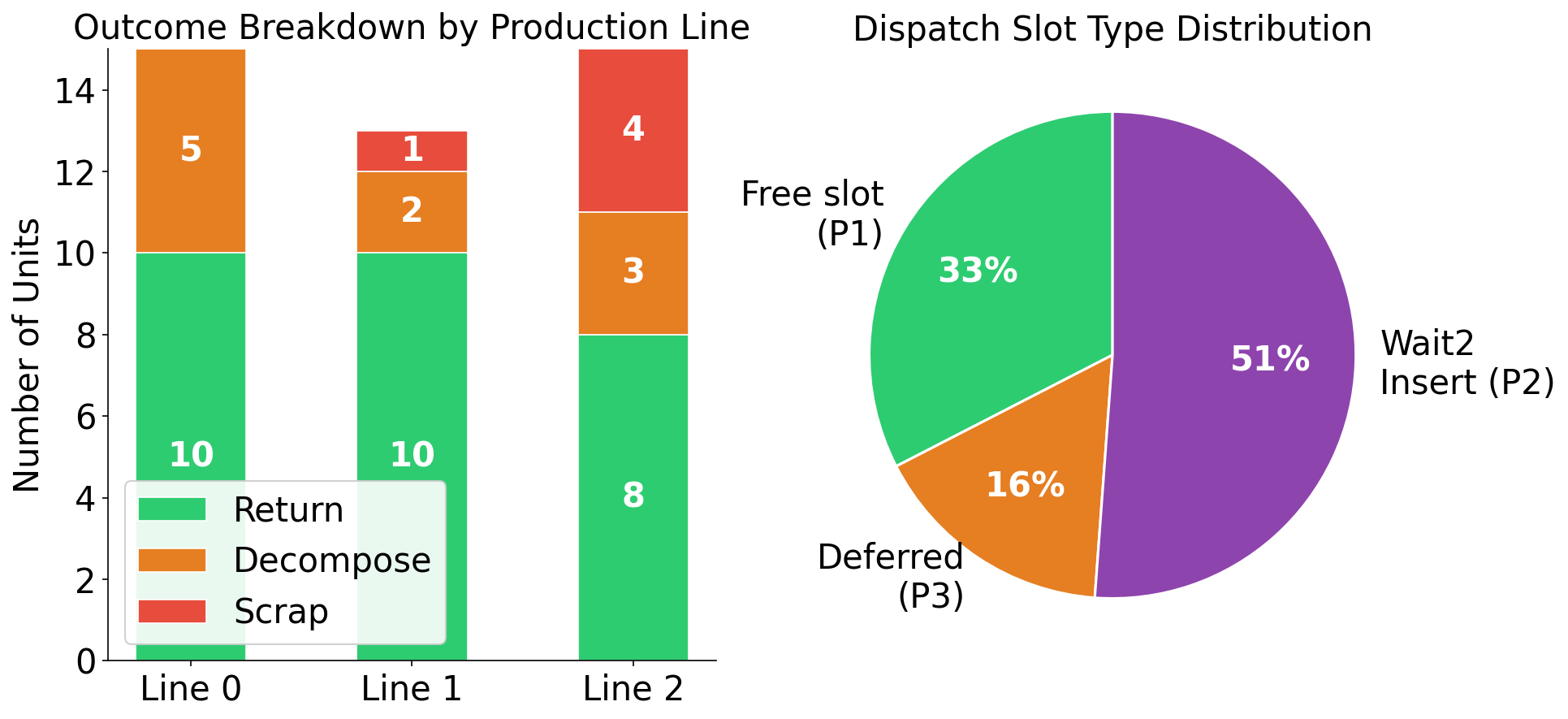}
  \caption{Left: outcome breakdown by production
           line. Right: proportion of units dispatched under
           each priority rule.}
  \label{fig:breakdown}
\end{figure}

Fig.~\ref{fig:breakdown} summarizes simulation outcomes at the aggregate level. The left panel confirms that \textsc{return} is the dominant outcome across all three production lines, consistent with $p_r = 0.90$ and $\Delta h = 0.80$ making full recovery the most likely trajectory. \textsc{decompose} and \textsc{scrap} are relatively rare and evenly distributed across lines, indicating no systematic line-level imbalance.

The right panel shows that 22 out of 43 units (51.2\%) were dispatched via the wait\,2 insertion rule (P2), with 14 units (32.6\%) dispatched to an immediately free line (P1) and 7 units (16.3\%) deferred (P3). The dominance of P2 reflects the frequent and prolonged wait\,2 windows generated by the repair loop, and confirms that the opportunistic insertion mechanism converts the majority of hold periods into active processing time, a throughput benefit that would be lost under a conventional first-come-first-served policy. 

\begin{comment}
\subsubsection{Summary}

Table~\ref{tab:results_summary} consolidates the key simulation metrics alongside the empirical benchmarks.

\begin{table}[ht]
\centering
\small
\caption{Simulation results ($M = 43$ units, $N = 3$ lines).}
\label{tab:results_summary}
\begin{tabular}{@{}lcc@{}}
\toprule
\textbf{Metric} & \textbf{Simulated} & \textbf{Empirical} \\
\midrule
Mean duration (days)          & 269.3 & 216.0 \\
Median duration (days)        & 234.0 & 170.0 \\
Max duration (days)           & 834.8 & 690.0 \\
Min duration (days)           &  14.7 &   1.0 \\
$W_1$ (days)                  & \multicolumn{2}{c}{53.3} \\
\midrule
\textsc{return}               & 30    & (69.8\%) \\
\textsc{decompose}            &  9    & (20.9\%) \\
\textsc{scrap}                &  4    & (9.3\%)  \\
\midrule
P1 (free slot)                & 14    & (32.6\%) \\
P2 (wait\,2 insert)           & 22    & (51.2\%) \\
P3 (deferred)                 &  7    & (16.3\%) \\
\bottomrule
\end{tabular}
\end{table}

\end{comment}
%%%%%%%%%%%%%%%%%%%%%%%%%%%%%%%%%%%%%%%%%%%%%%
\section{Conclusions}
\label{sec:conclusion}
%%%%%%%%%%%%%%%%%%%%%%%%%%%%%%%%%%%%%%%%%%%%%%

This paper has presented a Monte Carlo simulation framework for modeling the end-to-end repair logistics of PV inverter fleets at a centralized multi-line facility. The model captures the full unit lifecycle within an event-driven scheduler that exploits mandatory wait\,2 hold phases to insert additional units onto temporarily idle lines, improving throughput without adding physical capacity. Validated against field observations via the Wasserstein distance, the simulation successfully reproduces the bimodal structure of empirical repair durations, and the three-priority dispatch rule demonstrates that the majority of hold periods can be converted into productive processing time. 

Future work includes Bayesian joint calibration of the VaR and health parameters to further reduce the distributional gap, evaluation of alternative scheduling policies, such as health-score priority queuing, and coupling the model with a financial layer to quantify the revenue impact of reduced repair duration on plant availability.

% \section{References} 

\printbibliography
% \bibliography{reference}
\end{document}